\documentclass[slac_one]{revtex4}
\usepackage{graphicx}
\usepackage{fancyhdr}
\begin{document}

%Title of paper
\title{Leading Baryons and 
     {\boldmath $\sigma_{\rm tot}(\gamma p)$} at HERA} 
%% Paper title goes here

% Repeat the \author .. \affiliation  etc. as needed
%
% \affiliation command applies to all authors since the last
% \affiliation command. The \affiliation command should follow the
% other information

\author{W. B. Schmidke (on behalf of the H1 and ZEUS collaborations)}
\affiliation{Max Planck Institute for Physics, Munich, Germany}

\begin{abstract}
Leading baryon measurements from the H1 and ZEUS collaborations
are reported and compared to production models.
A new study of the energy dependence of the
photon-proton total cross section is also reported.
\end{abstract}

%\maketitle must follow title, authors, abstract
\maketitle

\thispagestyle{fancy}

% body of paper here - Use proper section commands
% References should be done using the \cite, \ref, and \label commands
% Put \label in argument of \section for cross-referencing
%\section{\label{}}

\section{Leading baryons}
\label{sec-lb}

Events with a baryon carrying a large fraction of the proton beam
energy have been observed in $ep$ scattering at 
HERA~\cite{lbh1,lbzeus}.
The dynamical mechanisms for their production are not completely
understood.
They may be the result of hadronization of the proton remnant,
leaving a baryon in the final state.
Exchange of virtual particles is also expected to contribute.
In this picture, the target proton fluctuates into a virtual
meson-baryon state.
The virtual meson scatters with the projectile lepton, leaving
the fast forward baryon in the final state.
Leading neutron (LN) production occurs through the exchange
of isovector particles, notably the $\pi^+$ meson.
For leading proton (LP) production isoscalar exchange
also contributes, including
diffraction mediated by Pomeron exchange.
In the exchange picture, the cross section for some process 
in $ep$ scattering with e.g. LN production factorizes:
\mbox{$
\sigma_{ep \rightarrow enX} = f_{\pi / p}(x_L,t) \cdot
                              \sigma_{e \pi \rightarrow eX}.
$}
Here $f_{\pi / p}$ is the flux of virtual pions in the proton,
$x_L = E_n/E_p$ is the fraction of the proton beam energy carried
by the neutron, and $t$ is the virtuality of the exchanged pion.
$\sigma_{e \pi \rightarrow eX}$ is the cross section for
electroproduction on the pion.

% FIG. 1 -----------------------------------------------------------------------
 \begin{figure}[h]
\begin{tabular}{cc}

 \begin{minipage}{8cm}
\includegraphics[width=7.5cm]{DESY-07-011_8_bins7-15.epsi}
 \end{minipage}
&
 \begin{minipage}{6cm}
\includegraphics[width=6.0cm]{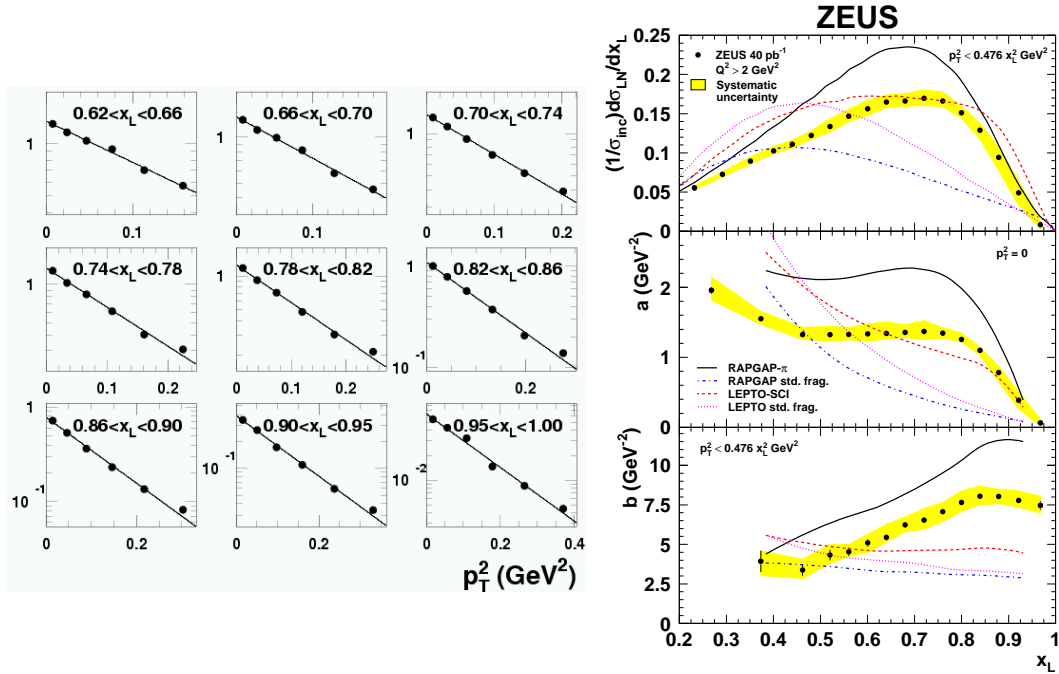}
 \end{minipage}

 \end{tabular}
\caption{
Left: LN $p_T^2$ distributions in bins of $x_L$
in the range $p_T^2 < 0.476 \, x_L^2$ GeV$^2$,
where $p_T$ is the LN transverse momentum.
The lines are the result of exponential fits.
Right: LN $x_L$, intercept and slope distributions
compared to models.
Results are from the ZEUS collaboration~\cite{lbzeus}.
}
\label{fig-lndata}
\end{figure}

\subsection{Leading baryon production and models}

The left side of Fig. \ref{fig-lndata} shows the LN
$p_T^2$ distributions in bins of $x_L$.
They are well described by exponentials; thus
the parameterization
$ (1/\sigma_{\rm inc}) d^2 \sigma / dx_L dp_T^2 \propto a(x_L) \exp(-b(x_L) p_T^2)$
fully characterizes the two dimensional distribution.
Here $\sigma_{\rm inc}$ is the inclusive cross section without an LN requirement.
The right side of Fig. \ref{fig-lndata} shows the LN $x_L$,
intercept $a$ and slope $b$ distributions compared
to several models. The standard fragmentation models
implemented in {\sc Rapgap} and {\sc Lepto}
and the {\sc Lepto} model with soft color interactions
do not describe the data.
The {\sc Rapgap} model mixing standard fragmentation and pion exchange
gives a better description of the shape of the $x_L$ distribution,
and also predicts the rise of the slopes with $x_L$,
although both with too high values.

% FIG. 2 -----------------------------------------------------------------------
 \begin{figure}[h]
\begin{tabular}{lr}

 \begin{minipage}{8cm}
\includegraphics[width=7.5cm]{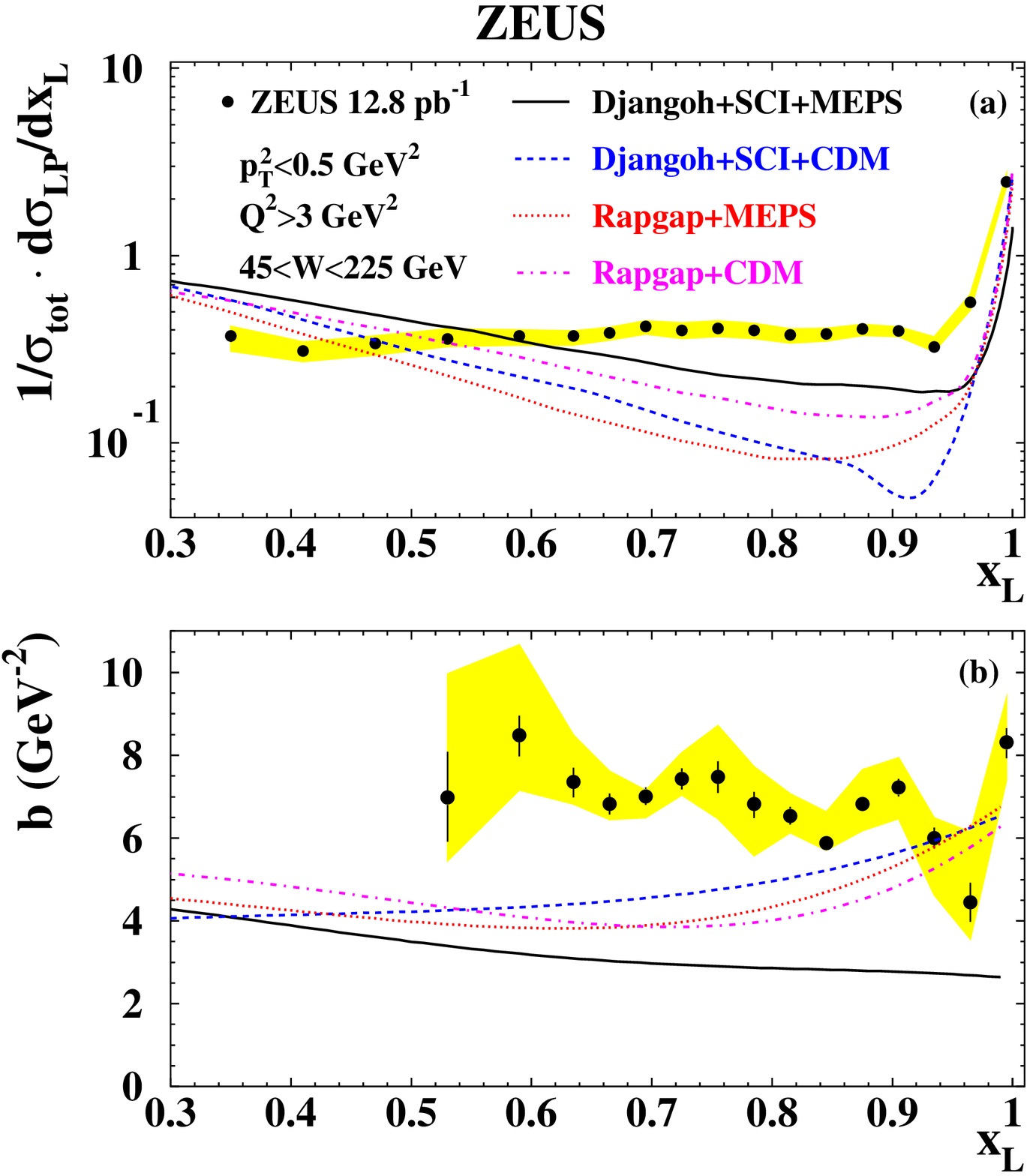}
 \end{minipage}
&
 \begin{minipage}{8cm}
\includegraphics[width=7.5cm]{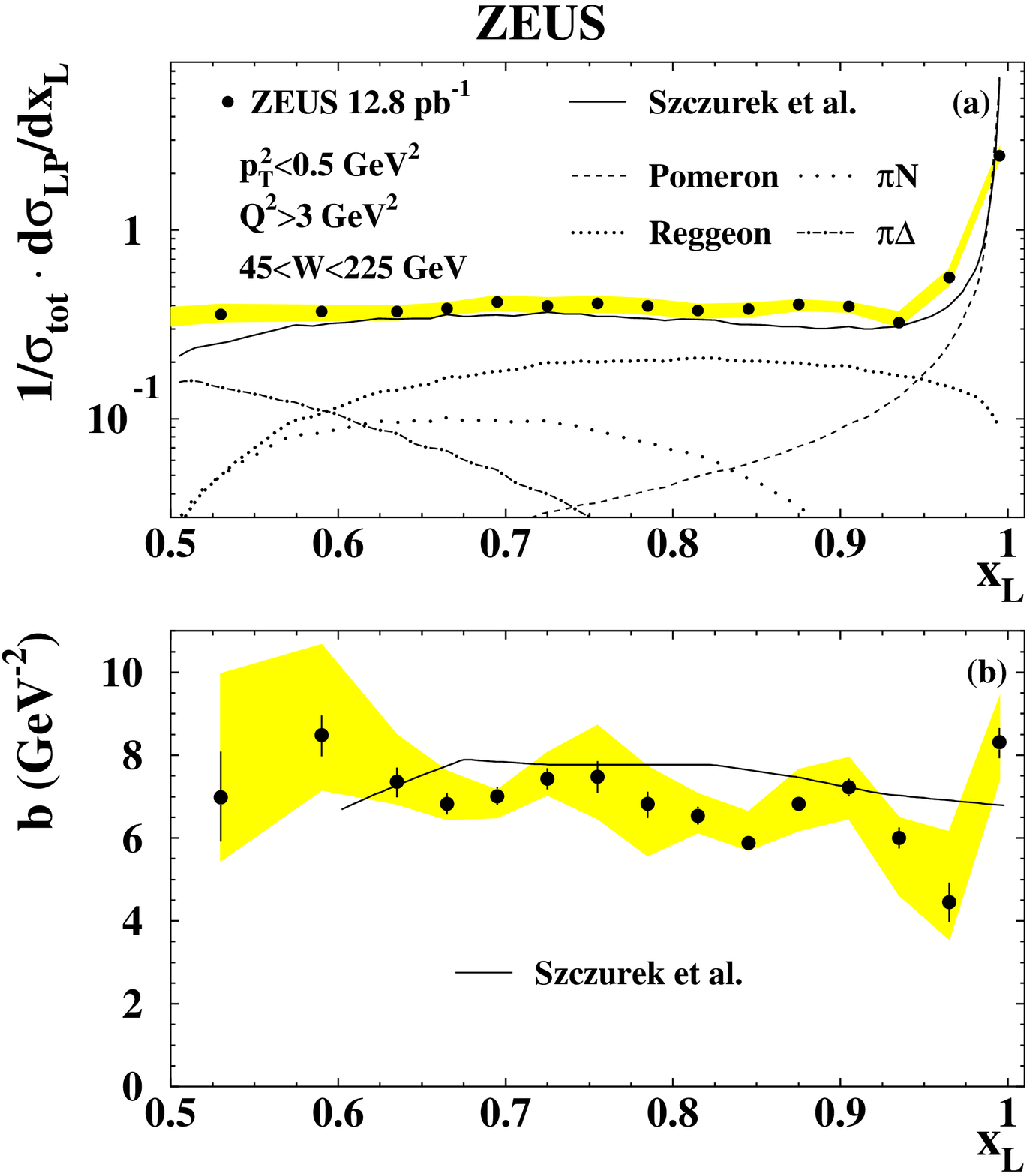}
 \end{minipage}

 \end{tabular}
\caption{
Left: LP $x_L$ distribution and exponential slopes compared
to standard fragmentation models.
Right: LP $x_L$ distribution and exponential slopes compared
to a model incorporating isoscalar and isovector exchanges.
Results are from the ZEUS collaboration~\cite{lbzeus}.
}
\label{fig-lpdata}
\end{figure}

If LP production proceeded only through isovector exchange, 
as LN production must, there should be half as many LP as LN.
The data (not shown) instead have approximately twice as many LP as LN.
Thus, exchanges of particles with isospins such as
isoscalars must be invoked for LP production.
The left side of Fig. \ref{fig-lpdata} shows a comparison of
the LP $x_L$ distributions and $p_T^2$ exponential slopes $b$ to
the {\sc Djangoh} and {\sc Rapgap}
Monte Carlo models incorporating standard fragmentation or
soft color interactions, none of which describe the data.
The right side of Fig. \ref{fig-lpdata} shows a comparison to
a model including exchange of both isovector and isoscalar
particles, including the Pomeron for diffraction~\cite{szczurek}.
These exchanges combine to give a good description of the
the $x_L$ distribution and slopes.

% FIG. 3 -----------------------------------------------------------------------
 \begin{figure}[h]
\begin{tabular}{cc}

 \begin{minipage}{8cm}
\includegraphics[width=7.5cm]{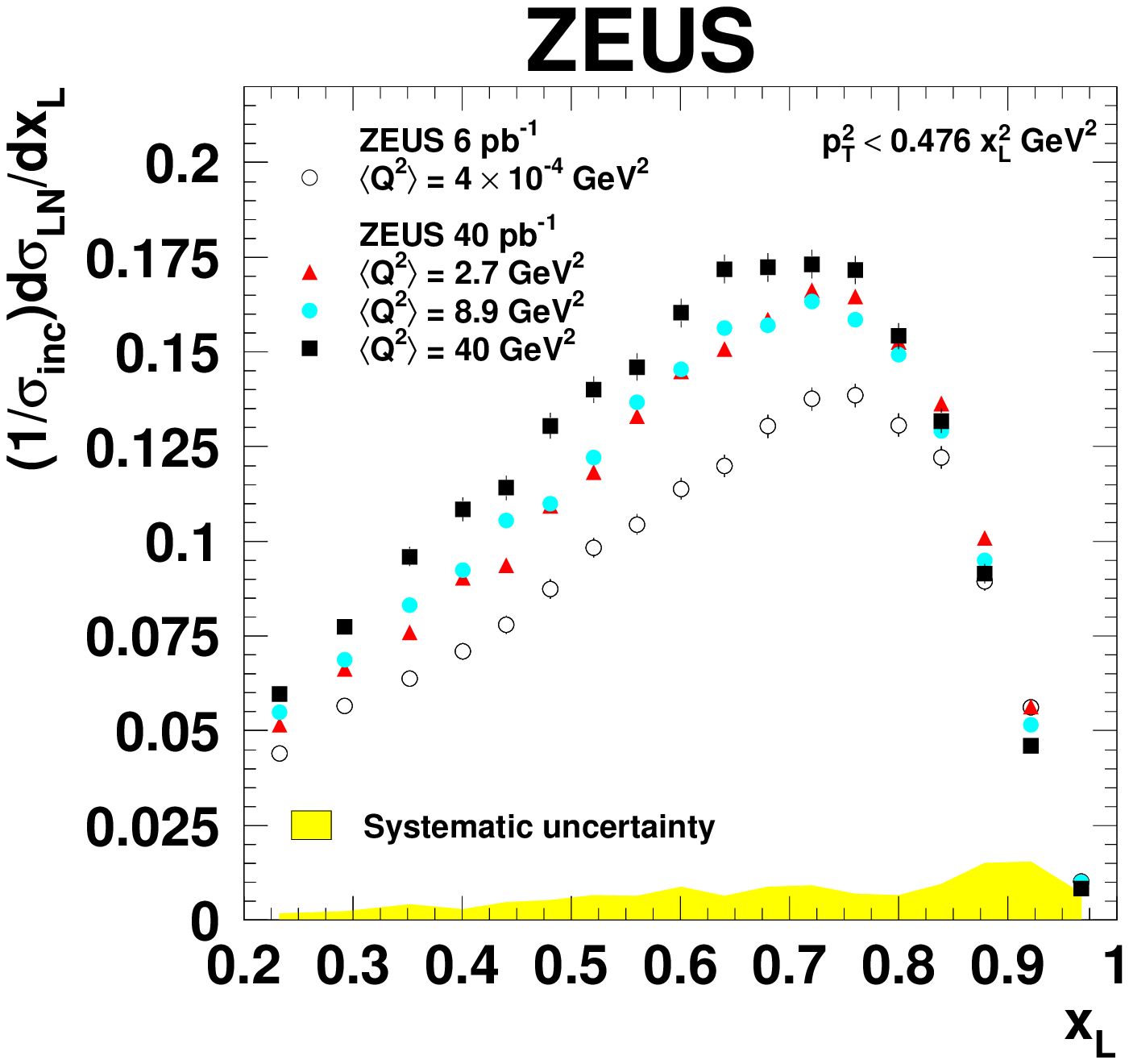}
 \end{minipage}
&
 \begin{minipage}{8cm}
\includegraphics[width=7.5cm]{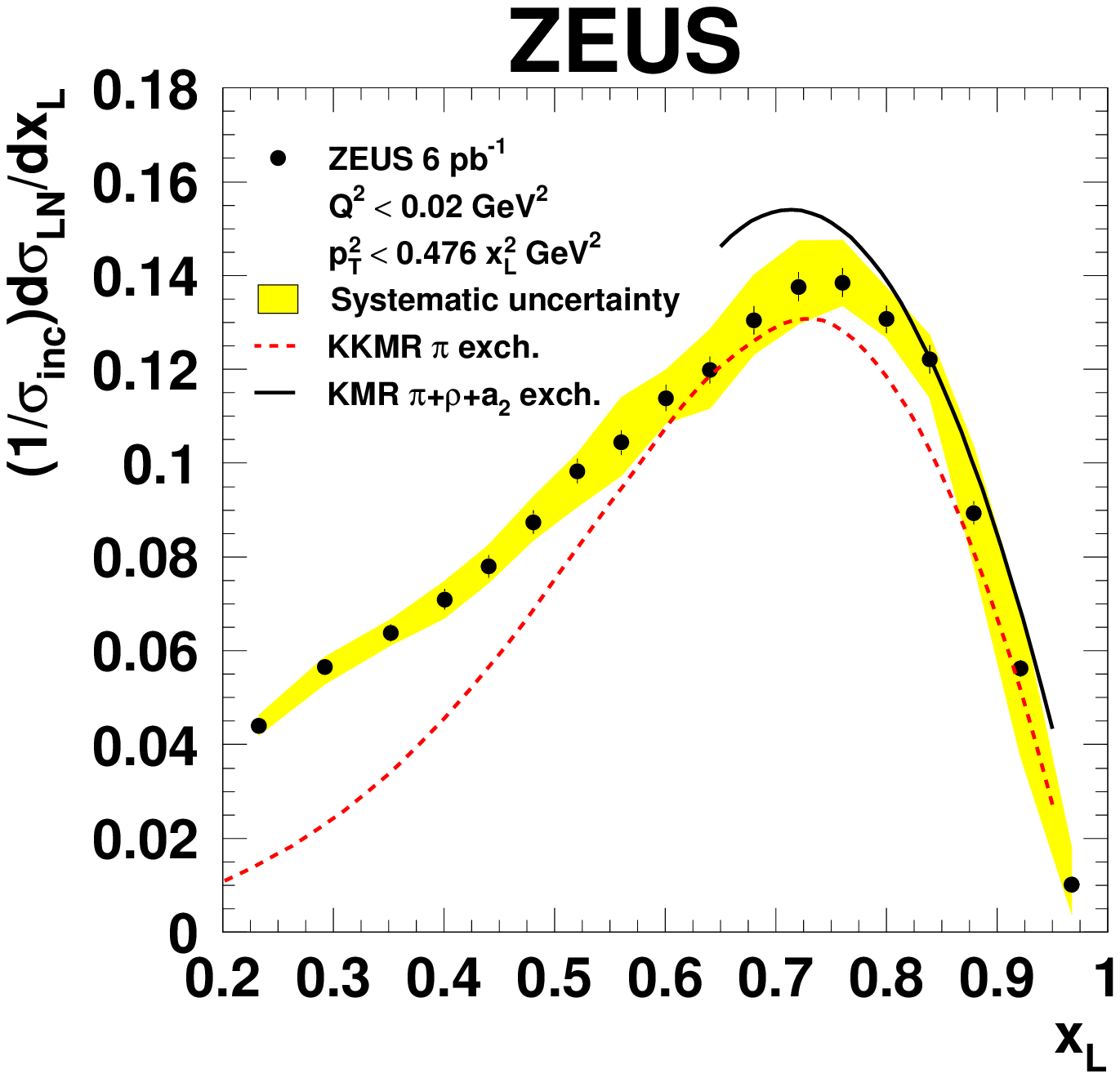}
 \end{minipage}

 \end{tabular}
\caption{
Left: LN $x_L$ distributions for photoproduction and
three bins of $Q^2$ in DIS.
Right: LN $x_L$ distributions for photoproduction compared
to exchange models including absorptive effects.
Results are from the ZEUS collaboration~\cite{lbzeus}.
}
\label{fig-absorp}
\end{figure}

\subsection{Absorption of leading neutrons}

The evidence for particle exchange in leading baryon production
motivates further investigation of the model.
One refinement of the simple picture described in the introduction
is absorption, or rescattering~\cite{abspubs}.
In this process, the virtual baryon also scatters with the
projectile lepton.
The baryon may migrate to lower $x_L$ or higher $p_T$ such that
it is outside of the detector acceptance, resulting in a
relative depletion of observed forward baryons.
The probability of this should increase with the size of
the exchanged photon.
The size of the photon is inversely related to its virtuality $Q^2$,
so the amount of absorption should increase with decreasing $Q^2$.

The left side of Fig. \ref{fig-absorp} shows the LN $x_L$ spectra
for photoproduction and for three bins of increasing $Q^2$.
The yield of LN increases monotonically with $Q^2$, in agreement with
the expectation of the decrease of loss through absorption
as $Q^2$ rises.
The right side of Fig. \ref{fig-absorp} shows the photoproduction data
with two predictions from models of meson exchange with absorption~\cite{kmr}.
The dashed curve model incorporates pion exchange with absorption,
accounting also for the migration in $x_L$ and $p_T$ of the neutron.
The solid curve model include the same effects, adding also
exchange of $\rho$ and $a_2$ mesons.
Both models give a good description of the large depletion of LN
in photoproduction relative to DIS seen in the left side of the figure.

% FIG. 4 -----------------------------------------------------------------------
 \begin{figure}[h]
\begin{tabular}{cc}

 \begin{minipage}{8cm}
\includegraphics[width=7.5cm]{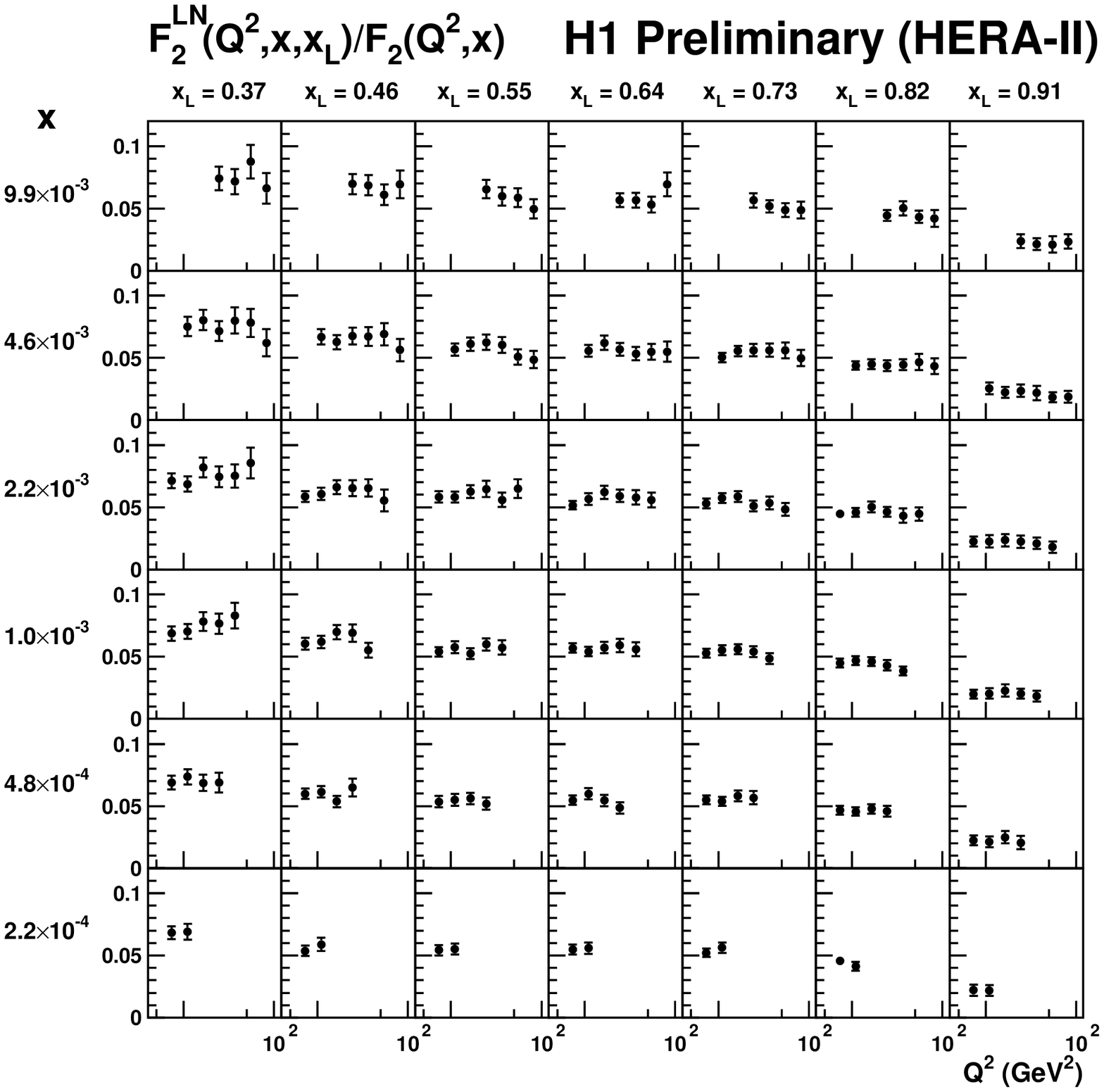}
 \end{minipage}
&
 \begin{minipage}{8.5cm}
\includegraphics[width=8.0cm]{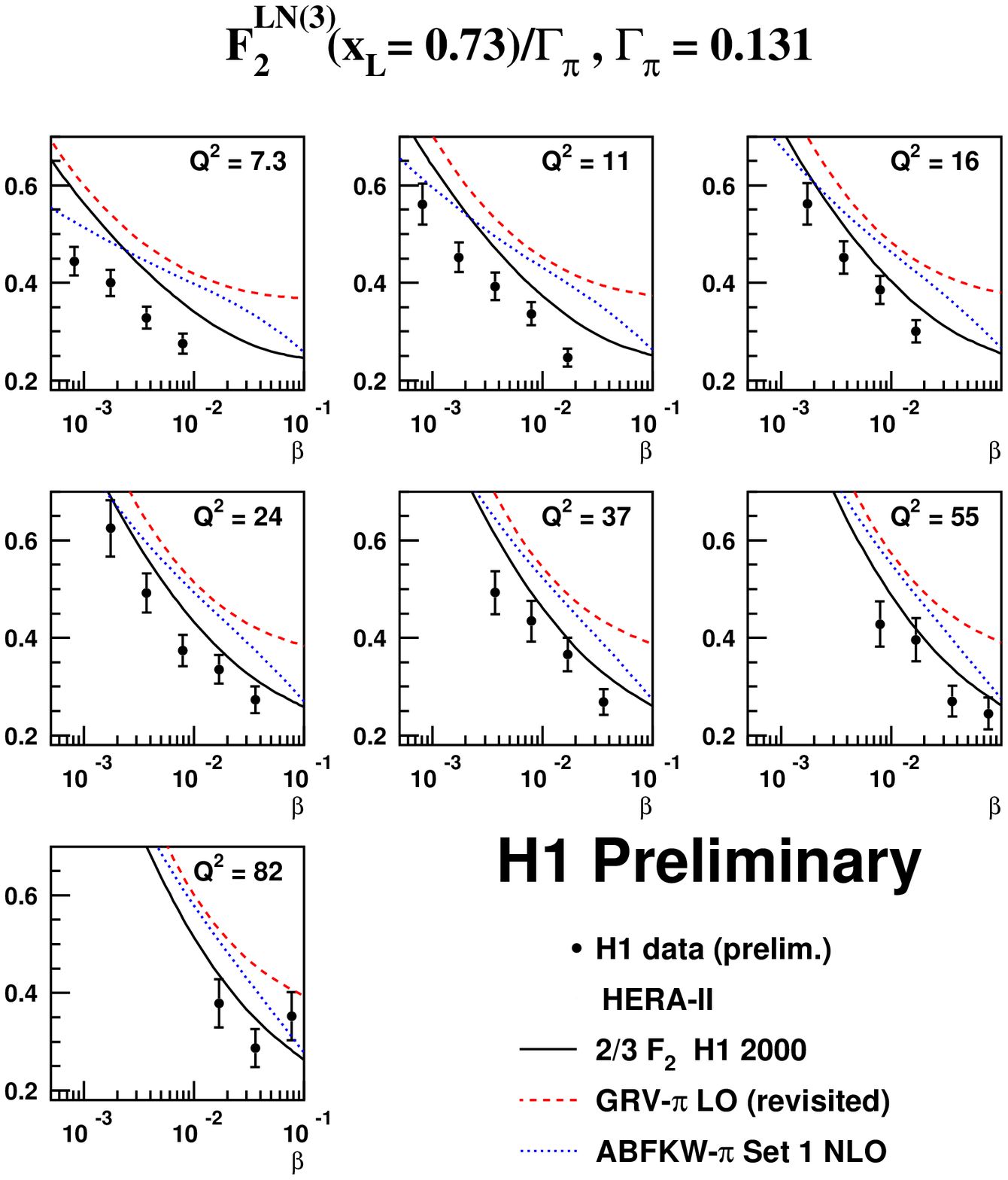}
 \end{minipage}

 \end{tabular}
\caption{
Left: Ratio of semi-inclusive LN to inclusive structure functions
as a function of $Q^2$ in bins of $x$ and $x_L$.
Right: Extracted pion structure function as a function of
$\beta = x/(1-x_L)$ in bins of $Q^2$.
The curves are the proton structure function scaled by 2/3
and two parameterizations based on Drell-Yan and direct
photon production data.
Results are from the H1 collaboration~\cite{lbh1}.
}
\label{fig-lnf2}
\end{figure}

\subsection{Pion structure function {\boldmath $F_2^{\pi}$}}

Analogous to the inclusive proton structure function $F_2(Q^2,x)$,
one can define an LN tagged semi-inclusive 
structure function $F_2^{LN(3)}(Q^2,x,x_L)$,
including also the dependence on $x_L$.
Here $x$ is the Bjorken scaling variable.
The left side of Fig. \ref{fig-lnf2} shows the ratios
$F_2^{LN}/F_2$ as a function of $Q^2$ in bins of $x$ and $x_L$.
Here $F_2^{LN}$ are the measured values from LN production in DIS
and the values of $F_2$ are obtained from
the H1-2000 parameterization~\cite{h1pdf}. 
For fixed $x_L$ the ratios are almost
flat for all $(x,Q^2)$ implying that $F_2^{LN}$ and $F_2$
have a similar $(x,Q^2)$ behavior. This result suggests the validity
of factorization, i.e. independence of the photon and the proton vertices.

The factorization relation can be rewritten
replacing the cross sections by $F_2^{LN} $ and $F_2^{\pi}$.
Using the measurement of $F_2^{LN(3)}$ for $0.68<x_L<0.77$,
and the integral over $t$ of the pion flux factor
at the center of this $x_L$ range,
$\Gamma_{\pi} = \int f_{\pi / p} \, dt = 0.131$,
one can extract the pion structure function
as $F_2^{\pi} = F_2^{LN(3)}/\Gamma_{\pi}$.
The right side of Fig. \ref{fig-lnf2} shows 
$F_2^{LN(3)}/\Gamma_{\pi}$ as a 
function of $\beta = x/(1-x_L)$ for fixed values of $Q^2$. 
The results are consistent with a previous ZEUS measurement~\cite{fpizeus}.
The data are compared to predictions of  parameterizations of the 
pion structure function~\cite{pipdfs},
and to the H1-2000 parameterization of the 
proton structure function~\cite{h1pdf} multiplied by the factor 2/3
according to naive expectation based on the number of valence quarks
in the pion and proton respectively.
The distributions show a steep rise 
with decreasing  $\beta$, in accordance with the pion and the proton
structure function parameterizations.
The scaled proton structure function gives the best description
of the data.

\section{Energy dependence of the photon-proton total cross section}

The energy dependences of hadronic total cross sections
can be described simply as the sum of two powers:
\mbox{
$\sigma_{\rm tot} = A \cdot W^{2 \epsilon} + B \cdot W^{-2 \eta}$}~\cite{dl},
where $W$ is the hadron-hadron center-of-mass energy.
The term with power $2 \epsilon$ is from Pomeron
exchange and is expected to be universal for all hadron-hadron reactions.
This has been studied at HERA in the $\gamma p$ total cross section,
where the photon fluctuates into a virtual hadron.
Previous HERA measurements had only one cross section
measurement at high $W$, and required results from
lower $W$ fixed-target experiments to extract $\epsilon$.

At the end of HERA running the proton beam energy was lowered
to half of its nominal value.
ZEUS took data for $\gamma p$ total cross section measurements
at both energies,
identifying photoproduction events with a positron tagger.
At these high values of $W$ the term with
power $2 \eta$ can be neglected, and $\epsilon$ can be
extracted from the ratio of $\sigma_{\rm tot}(\gamma p)$
at two energies.
By making the measurement with the same apparatus,
many acceptances and systematic effects in the ratio cancel.
The value extracted from the preliminary ZEUS measurement is
$\epsilon = 0.070 \pm 0.055$, consistent with the
value $\epsilon = 0.0808$ extracted from low-energy data~\cite{dl}.
The error on the ZEUS value will be reduced,
leading to an independent measurement of
the high energy dependence of hadronic total cross sections
with one apparatus.


\begin{thebibliography}{9}   % Use for  1-9  references
%\begin{thebibliography}{99} % Use for 10-99 references

\bibitem{lbh1}
H1 Coll., contribution to ICHEP-08, H1prelim-08-111.

\bibitem{lbzeus}
ZEUS Coll., S.~Chekanov et al., Nucl.~Phys. {\bf B~776}, 1 (2007),
and references cited therein.

\bibitem{szczurek}
A.~Szczurek, N.N.~Nikolaev and J.~Speth, Phys.~Lett. {\bf B~428}, 383 (1998).

\bibitem{abspubs}
N.N.~Nikolaev, J.~Speth and B.G.~Zakharov, 
Preprint KFA-IKP(TH)-1997-17 (hep-ph/9708290) (1997);
U.~D'Alesio and H.J.~Pirner, Eur.~Phys.~J. {\bf A~7}, 109 (2000).

\bibitem{kmr}
V.A.~Khoze, A.D.~Martin and M.G.~Ryskin, Eur.~Phys.~J. {\bf C~48},
  797 (2006).

\bibitem{h1pdf}
H1 Coll., C.~Adloff et al., Eur.~Phys.~J. {\bf C~21}, 33 (2001).

\bibitem{fpizeus}
ZEUS Coll., S.~Chekanov et al., Nucl.~Phys. {\bf B~637}, 3 (2002).

\bibitem{pipdfs}
P.~Aurenche et al., Phys.~Lett. {\bf B~233}, 517 (1989);
M.~Gl\"{u}ck, E.~Reya and I.~Schienbein, Eur.~Phys.~J. {\bf C~10},
 313 (1999).

\bibitem{dl}
A.~Donnachie and P.V.~Landshoff, Phys.~Lett. {\bf B~296}, 227 (1992).

\end{thebibliography}
\end{document}